
\input harvmac
\noblackbox
\newcount\figno
\figno=0
\def\fig#1#2#3{
\par\begingroup\parindent=0pt\leftskip=1cm\rightskip=1cm\parindent=0pt
\baselineskip=11pt
\global\advance\figno by 1
\midinsert
\epsfxsize=#3
\centerline{\epsfbox{#2}}
\vskip 12pt
{\bf Fig. \the\figno:} #1\par
\endinsert\endgroup\par
}
\def\figlabel#1{\xdef#1{\the\figno}}
\def\encadremath#1{\vbox{\hrule\hbox{\vrule\kern8pt\vbox{\kern8pt
\hbox{$\displaystyle #1$}\kern8pt}
\kern8pt\vrule}\hrule}}

\input epsf

\overfullrule=0pt


%

\def\tilde{\widetilde}

%
\def\inbar{\,\vrule height1.5ex width.4pt depth0pt}
\def\IB{\relax{\rm I\kern-.18em B}}
\def\IC{\relax\hbox{$\inbar\kern-.3em{\rm C}$}}
\def\ID{\relax{\rm I\kern-.18em D}}
\def\IE{\relax{\rm I\kern-.18em E}}
\def\IF{\relax{\rm I\kern-.18em F}}
\def\IG{\relax\hbox{$\inbar\kern-.3em{\rm G}$}}
\def\IH{\relax{\rm I\kern-.18em H}}
\def\II{\relax{\rm I\kern-.18em I}}
\def\IK{\relax{\rm I\kern-.18em K}}
\def\IL{\relax{\rm I\kern-.18em L}}
\def\IM{\relax{\rm I\kern-.18em M}}
\def\IN{\relax{\rm I\kern-.18em N}}
\def\IO{\relax\hbox{$\inbar\kern-.3em{\rm O}$}}
\def\IP{\relax{\rm I\kern-.18em P}}
\def\IQ{\relax\hbox{$\inbar\kern-.3em{\rm Q}$}}
\def\IR{\relax{\rm I\kern-.18em R}}
\font\cmss=cmss10 \font\cmsss=cmss10 at 7pt
\def\IZ{\relax\ifmmode\mathchoice
{\hbox{\cmss Z\kern-.4em Z}}{\hbox{\cmss Z\kern-.4em Z}}
{\lower.9pt\hbox{\cmsss Z\kern-.4em Z}}
{\lower1.2pt\hbox{\cmsss Z\kern-.4em Z}}\else{\cmss Z\kern-.4em
Z}\fi}
\def\IGa{\relax\hbox{${\rm I}\kern-.18em\Gamma$}}
\def\IPi{\relax\hbox{${\rm I}\kern-.18em\Pi$}}
\def\ITh{\relax\hbox{$\inbar\kern-.3em\Theta$}}
\def\IOm{\relax\hbox{$\inbar\kern-3.00pt\Omega$}}

\font\zfont = cmss10 

\def\bigone{\hbox{1\kern -.23em {\rm l}}}
\def\ZZ{\hbox{\zfont Z\kern-.4emZ}}

\def\a{\alpha}
\def\b{\beta}

\def\IR{\relax{\rm I\kern-.18em R}}
\def\I1{\relax{\rm I\kern-.6em 1}}
\def\Dsl{\,\raise.15ex\hbox{/}\mkern-13.5mu D}
\def\Gsl{\,\raise.15ex\hbox{/}\mkern-13.5mu G}
\def\Csl{\,\raise.15ex\hbox{/}\mkern-13.5mu C}
\font\cmss=cmss10 \font\cmsss=cmss10 at 7pt

\def\cp{{\cal P}}
\def\co{{\cal O}}

\font\zfont = cmss10 

\def\bigone{\hbox{1\kern -.23em {\rm l}}}
\def\ZZ{\hbox{\zfont Z\kern-.4emZ}}

\def\IL{\relax{\rm I\kern-.18em L}}
\def\IH{\relax{\rm I\kern-.18em H}}
\def\IR{\relax{\rm I\kern-.18em R}}
\def\IC{\relax\hbox{$\inbar\kern-.3em{\rm C}$}}
\def\IZ{\relax\ifmmode\mathchoice
{\hbox{\cmss Z\kern-.4em Z}}{\hbox{\cmss Z\kern-.4em Z}}
{\lower.9pt\hbox{\cmsss Z\kern-.4em Z}}
{\lower1.2pt\hbox{\cmsss Z\kern-.4em Z}}\else{\cmss Z\kern-.4em
Z}\fi}


\font\manual=manfnt \def\dbend{\lower3.5pt\hbox{\manual\char127}}

\def\IZ{\relax\ifmmode\mathchoice
{\hbox{\cmss Z\kern-.4em Z}}{\hbox{\cmss Z\kern-.4em Z}}
{\lower.9pt\hbox{\cmsss Z\kern-.4em Z}}
{\lower1.2pt\hbox{\cmsss Z\kern-.4em Z}}\else{\cmss Z\kern-.4em
Z}\fi}

\def\cc{{\cal C }}


\def\IZ{\relax\ifmmode\mathchoice
{\hbox{\cmss Z\kern-.4em Z}}{\hbox{\cmss Z\kern-.4em Z}}
{\lower.9pt\hbox{\cmsss Z\kern-.4em Z}}
{\lower1.2pt\hbox{\cmsss Z\kern-.4em Z}}\else{\cmss Z\kern-.4em
Z}\fi}
\def\IB{\relax{\rm I\kern-.18em B}}
\def\IC{{\relax\hbox{$\inbar\kern-.3em{\rm C}$}}}
\def\ID{\relax{\rm I\kern-.18em D}}
\def\IE{\relax{\rm I\kern-.18em E}}
\def\IF{\relax{\rm I\kern-.18em F}}
\def\IG{\relax\hbox{$\inbar\kern-.3em{\rm G}$}}
\def\IGa{\relax\hbox{${\rm I}\kern-.18em\Gamma$}}
\def\IH{\relax{\rm I\kern-.18em H}}
\def\II{\relax{\rm I\kern-.18em I}}
\def\IK{\relax{\rm I\kern-.18em K}}
\def\IP{\relax{\rm I\kern-.18em P}}

\def\IQ{\relax\hbox{$\inbar\kern-.3em{\rm Q}$}}
\def\IP{\relax{\rm I\kern-.18em P}}
\def\I1{\relax{\rm 1\kern-.18em I}}

\def\inbar{\,\vrule height1.5ex width.4pt depth0pt}

\font\cmss=cmss10 \font\cmsss=cmss10 at 7pt
\def\IR{\relax{\rm I\kern-.18em R}}


\def\boxit#1{\vbox{\hrule\hbox{\vrule\kern8pt
\vbox{\hbox{\kern8pt}\hbox{\vbox{#1}}\hbox{\kern8pt}}
\kern8pt\vrule}\hrule}}
\def\mathboxit#1{\vbox{\hrule\hbox{\vrule\kern8pt\vbox{\kern8pt
\hbox{$\displaystyle #1$}\kern8pt}\kern8pt\vrule}\hrule}}


\def\inbar{\,\vrule height1.5ex width.4pt depth0pt}

\font\cmss=cmss10 \font\cmsss=cmss10 at 7pt
\def\IR{\relax{\rm I\kern-.18em R}}

\lref\mns{M. Douglas and G. Moore, ``D-branes, Quivers, and ALE
Instantons,'' hep-th/9603167\semi G. Moore, N. Nekrasov and S.
Shatashvili, ``Integrating Over Higgs Branches,'' hep-th/9712241.}
\lref\cs{P.~Van Nieuwenhuizen,
``General Theory of Coset Manifolds and Antisymmetric Tensors Applied to
Kaluza-Klein Supergravity,'' in ``Supersymmetry and Supergravity '84'',
Proceedings of the Trieste Spring School 1984.}
\lref\cdf{L.~Castellani, R.~D' Auria and P.~Fre, ``Supergravity and
superstrings: A Geometric perspective.'' {\it World Scientific}
(1991).}
\lref\str{A.~Strominger,
``Massless black holes and conifolds in string theory,'' Nucl.
Phys. {\bf B451}, 96 (1995) hep-th/9504090.}
\lref\keha{A.~Kehagias,
``New type IIB vacua and their F theory interpretation,'' Phys.
Lett. {\bf B435}, 337 (1998) hep-th/9805131.}
\lref\kw{I.R.~Klebanov and E.~Witten,
``Superconformal field theory on three-branes at a Calabi-Yau
singularity,'' Nucl. Phys. {\bf B536}, 199 (1998) hep-th/9807080.}
\lref\afhs{B.S.~Acharya, J.M.~Figueroa-O'Farrill, C.M.~Hull and B.~Spence,
``Branes at conical singularities and holography,''
hep-th/9808014.}
\lref\mp{D.R.~Morrison and M.R.~Plesser,
``Nonspherical horizons. 1,'' hep-th/9810201.}
\lref\co{P.~Candelas and X.C.~de la Ossa,
``Comments On Conifolds,'' Nucl. Phys. {\bf B342}, 246 (1990).}
\lref\pp{D.N.~Page and C.N.~Pope,
``Which Compactifications Of D = 11 Supergravity Are Stable?,''
Phys. Lett. {\bf 144B}, 346 (1984).}
\lref\jm{J.~Maldacena,
``The Large N limit of superconformal field theories and
supergravity,'' Adv. Theor. Math. Phys. {\bf 2}, 231 (1998)
hep-th/9711200.}
\lref\jdb{J.~de Boer,
``Six-dimensional supergravity on $S^3 \times AdS_3$ and 2-D
conformal field
                  theory,''
Nucl. Phys. {\bf B548}, 139 (1999) hep-th/9806104.}
\lref\msw{J.~Maldacena, A.~Strominger and E.~Witten,
``Black hole entropy in M theory,'' JHEP {\bf 12}, 002 (1997)
hep-th/9711053.}
\lref\mmt{R.~Minasian, G.~Moore and D.~Tsimpis,
``Calabi-Yau black holes and (0,4) sigma models,'' hep-th/9904217.}
\lref\bh{K.~Behrndt, G.~Lopes Cardoso, B.~de Wit, R.~Kallosh, D.~Lust and
T.~Mohaupt,
``Classical and quantum N=2 supersymmetric black holes,'' Nucl.
Phys. {\bf B488}, 236 (1997) hep-th/9610105.}
\lref\fks{S.~Ferrara, R.~Kallosh and A.~Strominger,
``N=2 extremal black holes,'' Phys. Rev. {\bf D52}, 5412 (1995)
hep-th/9508072.}
\lref\aa{G.~Moore,
``Arithmetic and attractors,'' hep-th/9807087; ``Attractors and
arithmetic,'' hep-th/9807056.}
\lref\hs{G.T.~Horowitz and A.~Strominger,
``Black strings and P-branes,'' Nucl. Phys. {\bf B360}, 197
(1991).}
\lref\dl{M.J.~Duff and J.X.~Lu,
``The Selfdual type IIB superthreebrane,'' Phys. Lett. {\bf B273},
409 (1991).}
\lref\wittkk{E.~Witten,
``Search For A Realistic Kaluza-Klein Theory,'' Nucl.\ Phys.\ {\bf
B186}, 412 (1981).}
\lref\crw{L.~Castellani, L.J.~Romans and N.P.~Warner,
``A Classification Of Compactifying Solutions For D = 11
Supergravity,'' Nucl.\ Phys.\ {\bf B241}, 429 (1984).}
\lref\dpn{M.J.~Duff, B.E.~Nilsson and C.N.~Pope,
``Kaluza-Klein Supergravity,'' Phys.\ Rept.\ {\bf 130}, 1 (1986).}
\lref\dlp{M.J.~Duff, H.~Lu and C.N.~Pope,
``Supersymmetry without supersymmetry,'' Phys.\ Lett.\ {\bf B409},
136 (1997) hep-th/9704186; ``AdS(5) x S(5) untwisted,'' Nucl.\
Phys.\ {\bf B532}, 181 (1998) hep-th/9803061.}
\lref\wbh{E.~Witten,
``On string theory and black holes,'' Phys.\ Rev.\ {\bf D44}, 314
(1991).}
\lref\dkl{M.J.~Duff, R.R.~Khuri and J.X.~Lu,
``String solitons,'' Phys.\ Rept.\ {\bf 259}, 213 (1995)
hep-th/9412184.}
\lref\dwn{B.~de Wit and H.~Nicolai,
``A New SO(7) Invariant Solution Of D = 11 Supergravity,'' Phys.\
Lett.\ {\bf 148B}, 60 (1984).}
\lref\vnw{P.~van Nieuwenhuizen and N.P.~Warner,
``New Compactifications Of Ten-Dimensional And Eleven-Dimensional
Supergravity On Manifolds Which Are Not Direct Products,'' Commun.\
Math.\ Phys.\ {\bf 99}, 141 (1985).}

%

\Title{\vbox{\baselineskip12pt
\hbox{YCTP-P31-99}
\hbox{hep-th/9911042}
}} {\vbox{\centerline{ On the geometry of non-trivially embedded branes} }}

\bigskip
\centerline{Ruben Minasian$^{a,b}$ and Dimitrios Tsimpis$^a$}
\bigskip
\centerline{$^a$Department of Physics, Yale University}
\centerline{New Haven, CT 06520, USA}

\bigskip
\centerline{$^b$Centre de Physique Th\'eorique
}
\centerline{Ecole Polytechnique, F-91128 Palaiseau,  France }

\bigskip
\centerline{\bf Abstract}
\medskip
We present a formal supersymmetric solution of type IIB supergravity
generalizing previously known solutions corresponding to D3 branes to geometries
without an orthogonal split between parallel and transverse directions. The
metric is given implicitly as one with respect to which a certain connection is
compatible. The case of the deformed conifold is discussed in detail.

\Date{November 5, 1999}

%

\newsec{Introduction and motivation}

The $AdS$/CFT correspondence in cases with lower supersymmetries involving the
two- and three-dimensional  $AdS$ spaces is much less understood than the others
\jm. In particular, the near-horizon geometry of black strings in simple
five-dimensional supergravity (obtained by compactification of M-theory on a
Calabi-Yau threefold $X$) is given by $AdS_2 \times S^2.$ Here we meet a little
puzzle discussed also in \jdb. The spectrum of the supergravity theory is
defined by $H^2(X, \IZ),$ and when considering a magnetic string solution, one
naturally writes down a string coupled to $(h_{11}(X) -1)$ vectors and a
graviphoton. However as pointed out in \msw, the theory on the string is
governed by a much larger lattice defined by $H^2(\cp, \IZ),$ where $\cp$ is the
four-cycle around which the M-theory fivebrane is wrapped. In particular, this
accounts for a very large entropy. This information about the cycle is not
reflected in the KK spectrum though. As seen in \mmt, the target space of the
dual $(0,4)$ conformal field theory is factorized. The coupling to supergravity
is governed mostly by the so called universal sector, while the numerous modes
making up the theory on the two-dimensional worldsheet are in the entropic
sector. Apparently, the usual procedure of compactifying M-theory and then
looking for the solutions misses the deformations of the cycle. In doing so we
simply find a solution corresponding to the universal sector\foot{The full
entropy is recovered by loop corrections to supergravity \bh.}, and
schematically we can write
$$[{\rm compactification}, {\rm solution}] = {\rm entropic \,\,\,\,
sector.}$$
Finding a complete solution with all the modes  seems to be extremely
hard.
Instead, we try to give a description of the 10d geometry corresponding to
the
situation in
\str: a D-brane wrapping a non-trivial cycle in an internal Calabi-Yau
space, shrinking to zero-size at certain points of the moduli space thus turning
into a massless BPS black hole from the four dimensional non-compact space point
of view\foot{Note that $N$ D3-branes do not form a bound state on the conifold.
To ensure the reliability of supergravity we are tacitly assuming that $N$ is
large.}.

D3 branes on conifold points have been discussed extensively in the $AdS$
literature recently \refs{\keha, \kw, \afhs, \mp}. However only the situation
where the branes are transverse to a CY manifold, and can be thought of as
spacetime-filling, has been addressed there. We will be concerned with a rather
different geometrical setup, the case of a D3 wrapping a 3-cycle in an internal
CY threefold.

In the next two sections, we will outline a procedure for finding formal
supersymmetric solutions under some general assumptions about the metric in the
absence of D3. Unfortunately, the metric of the solution is only given
implicitly as one with respect to which a certain connection is compatible.

As a specific case, we will consider type $IIB$ supergravity on a non-compact
Calabi-Yau threefold which we will call the {\it deformed conifold} $\cc
(\varepsilon)$. In doing this we are able to avoid the question of extra modes
since in our limiting case the three-fold is approximated by $T^*S^3$ and $S^3$
is rigid. The deformed conifold, which will be described in the following in
some detail, is topologically a 6d cone over an $S^2 \times S^3$ base, whose
apex is replaced by an $S^3$.

Before turning to the concrete example of D3 on a shrinking $S^3$ cycle in
section 5, we discuss the geometry of the singular and deformed conifolds of
\co. Using the machinery of coset-space geometry \refs{\cs, \cdf}, we are
able to give an explicit form of the deformed conifold metric presented
implicitly in
\co. A possible generalization to M2 and M5 cases is discussed in section
6.

\newsec{Branes on cycles}

We present a method for constructing formal supersymmetric solutions of type
$IIB$ supergravity corresponding to D3 branes, which generalizes previously
known solutions to geometries without an orthogonal split between directions
parallel and transverse to the brane. In particular, we do not assume that the
D3 is spacetime-filling but our method covers this situation as a special case.

In the presence of the D3 the 10d geometry will get deformed to account for the
back reaction due to the brane. This back reaction is captured by the {\it
warping factor}, a function of the coordinates transverse to the brane. There is
considerable amount of literature on the subject of supergravity solutions
corresponding to branes (see e.g. \refs{\hs, \dkl}). These solutions assume that
in the absence of D3 there is an orthogonal split of the 10d metric along
parallel and transverse to the brane directions:
\eqn\split{ds^2= \eta_{\mu \nu}dx^{\mu} dx^{\nu} + g_{mn}(y)dy^m dy^n}
where $\eta_{\mu \nu}$ is the (flat) metric on the cycle around which the D3
will wrap (which in that case is the whole 4d Minkowski spacetime) and
$g_{mn}(y)$ is the transverse metric. In the presence of the D3, the geometry is
modified
\eqn\intpod{ds^2= \Delta_{\perp}(y)\eta_{\mu \nu}dx^{\mu} dx^{\nu}
+ \Delta_{\vert \vert}(y)g_{mn}(y)dy^m dy^n  }
where $\Delta_{\perp}(y), \,\, \Delta_{\vert \vert}(y)$ are the warping
factors
(which turn out to be related) and are functions only of the coordinates
transverse to the brane.

Here we will not assume that there is the ``nice'' split of the form
\split.
Instead we will take the 10d metric in the absence of D3 to be of the form
\eqn\mtritapd{  ds^2= g_{\mu \nu}(x, \theta)dx^{\mu} dx^{\nu}+ f^2(U)dU^2
+ \sum_{a=1}^5{(e(x,\theta,U)^{a}{}_{\mu}dx^{\mu} +
e(x,\theta,U)^{a}{}_id\theta^i)^2} }
where $\{ x^{\mu}; \, \mu=0, \dots 3  \}$ are the coordinates parametrizing a
non-trivially embedded cycle $\cc$, $U$ is a ``radial'' coordinate in the
transverse space and $\{\theta^i; \, i, \dots 5  \}$ are ``angular'' coordinates
in  the transverse space. The metric on $\cc$ is $g_{\mu \nu}$ and it does not
depend on $U$. Note that $e(x,\theta,U)^{a}{}_{\mu}$ encodes the deviations from
orthogonally-split geometries. We will further assume that
\eqn\jhgd{e(x,\theta,U=0)^{a}{}_{\mu}=
e(x,\theta,U=0)^{a}{}_i=0,}
so that $\cc$ is at $U=0$. If we wish, we may consider $U$ as a collective label
for a set of ``radial '' coordinates $U_1, U_2,\dots$ which enter metric
\mtritapd\ as $f^2_1(U_1)dU_1^2+f^2_2(U_2)dU_2^2+\dots$

For simplicity we will drop the $\theta$ dependence of $g_{\mu
\nu}$ in the following. All our arguments of section 3
go through for $\theta$-dependent $g_{\mu \nu}$ as well. Let us remark that if
we keep the $\theta$ dependence, it appears that in the flat D3 limit we may
recover {\it warped} $AdS_5
\times {}_wS^5$ products. The possibility of such supergravity vacua
has been known for some time \refs{\dwn, \vnw}, however these haven't
appeared
as  brane near-horizon limits so far. This discussion may provide a brane
realization of such vacua.

The metric of \mtritapd\ is of some generality. A trivial example would be the
case where the 10d metric is that of a direct sum of 4d Minkowski plus a 6d
cone, with $U$ being the distance (in the ten-dimensional sense) from the apex.
Another example, which will be discussed in the following, is the geometry of
the deformed conifold near the $S^3$ at the apex.

\newsec{The solution}

In this section we will discuss how does \mtritapd\ change in the presence of a
D3 along $\cc$. As already emphasized, the solution will be given implicitly:
the metric in the presence of D3 is the metric with respect to which the
connection of equation (3.8) below, is metric compatible. This statement makes
sense since for every torsion-free connection with $SO(N)$ holonomy (where $N$
is the dimension of the manifold) there is an essentially unique metric with
respect to which the connection is compatible. We can see this as follows: take
the metric at a given point to be some constant symmetric $N \times N$ matrix.
Parallel-transport it using the connection to define the metric at any other
point. The absence of inconsistencies under parallel transport along closed
loops is equivalent to the requirement of $SO(N)$ holonomy. The metric
constructed in this way is defined up to rigid $GL(N,\IR)$ coordinate
transformations.

We want to warn the reader that as we do not have the charge distribution
explicitly, nothing excludes the possibility that this solution corresponds to a
completely ``smeared'' D3. In that case it would be wrong to think of the D3 as
wrapping $\cc$. It is more correct to say that $\cc$ will be identified with the
horizon in the presence of the D3\foot{We would like to thank G. Moore for
explaining this to us.}.

\subsec{Conventions}

$\mu, \, \nu$ are curved indices for the directions along the D3.

$m, \, n$ are curved indices for the coordinates transverse to the D3 including
U.

$i, \, j$ are curved indices for the coordinates transverse to the D3 excluding
U.

$M, \, N$ are ten-dimensional curved indices.

$\a, \, \b$ are flat indices for the directions along the D3.

$a, \, b$ are flat indices corresponding to the directions transverse to the D3
excluding U.

$\bullet$ is the flat index corresponding to $U$.

$A, \, B$ are ten-dimensional flat indices.

\subsec{The solution}

It will be useful to give the coframe version of \mtritapd:
\eqn\cfv{ds^2= \sum_{\a=0}^3{(e^{\a})^2}+ \sum_{a=1}^5{(e^{a})^2}
+(e^{\bullet})^2  }
where
\eqn\where{\eqalign{e^{\a}=&e(x)^{\a}{}_{\mu}dx^{\mu}; \,\,\,\,
e^{\bullet}=e(U)^{\bullet}{}_{U}dU= f(U)dU \cr
e&^{a}=e(x,\theta,U)^{a}{}_{\mu}dx^{\mu} + e(x,\theta,U)^{a}{}_id\theta^i \cr}}
The fields of $IIB$ are the graviton $g_{MN}$, a complex scalar $\tau$
parametrizing an $SL(2, \IR)/U(1)$ coset space, a pair of two-forms
$B^{1,2}_{MN}$ which form an $SL(2, \IR)$ doublet, a self-dual four-form
$A^{(4)}$ with field strength $F^{(5)}$, and two complex-Weyl fermions: a
gravitino $\psi_{M}$ and a dilatino $\lambda$.

The supersymmetry transformations are parametrized by a complex-Weyl
spinor
$\epsilon$ and, in a background with all fields set to zero except for the
graviton and the four-form, only the gravitino transformation is not
identically
zero
\eqn\gravtrnsfm{\delta_{\epsilon}\psi_A=D(\Omega)_A\epsilon +
{i \over 4\times5!}\Gamma^{A_1\dots A_5}F_{A_1\dots A_5} \Gamma_A \epsilon
}
where
\eqn\covdev{D(\Omega)_A=\partial_A +
{1 \over 4}\Omega_A{}^{BC}\Gamma_{BC}; \,\,\,\,
\partial_A=e_A{}^M\partial_M  }
and $\Omega^A{}_B$ is the connection one-form corresponding to the 10d metric in
the presence of the D3.

Our ansatz for the five-form is given in terms of one function of $U$,
$C(U)$
\eqn\ansfff{ F_{\a_0\a_1\a_2\a_3
\bullet}=\varepsilon_{\a_0\a_1\a_2\a_3}C(U);
\,\,\,\, F_{a_1a_2a_3a_4a_5}= \varepsilon_{a_1a_2a_3a_4a_5}{}^{
\bullet}C(U)  }
with all other components equal to zero. We then find
\eqn\wthfnd{\eqalign{\Gamma^{A_1\dots A_5}F_{A_1\dots A_5} \Gamma_{\a}
&= -i5!C(U)\Gamma_{\bullet}\Gamma_{\a}(\rho^{(6)}+\rho^{(4)} )  \cr
\Gamma^{A_1\dots A_5}F_{A_1\dots A_5} \Gamma_{a}
&= i5!C(U)\Gamma_{\bullet}\Gamma_{a}(\rho^{(6)}+\rho^{(4)}) \cr}}
where
\eqn\chrls{\rho^{(4)}:=i\Gamma^{\a=0} \dots \Gamma^{\a=3};
\,\,\,\, \rho^{(6)}:= -i\Gamma^{a=1} \dots \Gamma^{a=5}  }
are the ``parallel'' and ``transverse'' chirality operators. Let $\omega^A{}_B$
be the connection one-form associated to the metric \cfv. Our ansatz for the 10d
metric in the presence of the D3 will be given implicitly by requiring that it
be associated to the connection $\Omega^A{}_B$ given by
\eqn\cnanst{ \Omega^A{}_B=\omega^A{}_B + e^A \partial_B ln\Delta(U)_{(A)}
-
e_B \partial^A ln\Delta(U)_{(B)}}
where
\eqn\defofd{\Delta(U)_{(A)}= \cases{\Delta(U)_{\vert \vert} \,\,\,\, ,A=\a
\cr
\Delta(U)_{\perp} \,\,\,\, ,A=a  } }
are two warping factors.

In \keha\ a relation identical to \cnanst\ holds, arising from a rescaling of
the vielbeins $e^A
\rightarrow \Delta_{(A)} e^A$. Our case however is very
different as it does {\it not} imply that the associated metrics are
related by
such a rescaling. The reason is that, as follows from \cfv, \where, the
connection $\omega^A{}_B$ generally has non zero components of the form
$\omega^\a{}_a$ mixing transverse with parallel directions.

We will now assume that
\eqn\chrlty{\rho^{(4)}\epsilon= \rho^{(6)}\epsilon= \epsilon }
the susy transformations then read:
\eqn\susytrnsfmsthr{\eqalign{0&= [D(\omega)_{\a} +
\Gamma_{\a \bullet}(\partial_U ln\Delta_{\vert \vert}+{1 \over 2}C
)]\epsilon \cr
 0 &=[D(\omega)_{a} -{1 \over 2}C\delta_{a \bullet}+
\Gamma_{a \bullet}(\partial_U ln\Delta_{\perp}-{1 \over 2}C  )]\epsilon
\cr }
}
Setting
\eqn\ssyanscc{\eqalign{C(U)&= 2\partial_U ln\Delta(U)_{\perp}  \cr
\Delta(U)_{\perp} &= {1 \over \Delta(U)_{\vert \vert}} \cr
\epsilon &= \Delta(U)_{\vert \vert} \hat{\epsilon}   \cr  }  }
equation \susytrnsfmsthr\ reduces to
\eqn\fnssstr{D(\omega)_A \hat{\epsilon} =0,}
i.e. the solution preserves some supersymmetry provided the geometry
\cfv\ in the absence of D3 admits a covariantly constant spinor. The
integrability of
\fnssstr\ is equivalent to the requirement of Ricci-flatness for the
geometry in the absence of D3.
\eqn\rrfl{Ric(\omega)_{AB}=0}
To check the consistency of our ansatz with Einstein equations
\eqn\eins{Ric(\Omega)_{AB}={1 \over 4\times 4!} F_A{}^{A_1 \dots A_4}F_{BA_1 \dots A_4}}
we simply substitute \cnanst\ into \eins\ taking \rrfl\ into
account. The result is \refs{\dl, \keha}:
\eqn\harmcnd{D(\omega)_U D(\omega)^U  \Delta_{\perp}(U)^2=0}
This is the condition that $ \Delta_{\perp}(U)^2$ is harmonic. In proving
the
above we used the fact that
\eqn\offdvn{\omega^{\bullet}{}_\a =0 }
We can see this as follows. From \cfv\ we get
\eqn\dgfhj{0=d(e(U)^{\bullet}{}_U dU )=de^{\bullet}=-\omega^{\bullet}{}_\a
e^\a -\omega^{\bullet}{}_a e^a}
therefore the only possibly nonzero components of $\omega^{\bullet}{}_\a$
are of
the form $\omega_\a{}^{\bullet}{}_\a$. On the other hand
\eqn\oiu{de^\a= -\omega^\a{}_{\bullet} e^{\bullet}+ \dots=
-\omega_\a{}^\a{}_{\bullet}e^\a \wedge e^{\bullet}+ \dots }
But since $\partial_U e^\a =0$, $de^{\a}$ cannot have a piece proportional
to
$dU$ and we conclude that $\omega^{\bullet}{}_\a =0$.

\newsec{Geometry of conifolds}

The main purpose of this section is to provide a specific example of the
geometrical setup under which our solution  of the previous section is valid.
Indeed we show (see subsec. 4.5) that the near-horizon limit of the deformed
conifold is a particular case to which our method applies.

None of this section is new. The results are in principle contained in previous
works \refs{ \co, \cs, \cdf}. However we want to draw the attention of the
reader to two points. Equation (4.15) below, contains a term (the one
proportional to $B$) which is usually omitted from discussions in the literature
related to $T^{1,1}$ spaces. However this term appears naturally in the metric
of the deformed conifold, explicitly presented in equation (4.35). In section 5
of \co\ the metric is given implicitly in terms of 7 variables (one radius and 6
Euler angles) in a form which makes it difficult to distinguish 6 independent
ones. For these reasons we think the discussion of this section is useful.

\subsec{$T^{p,q}$ spaces}

A cone $\cc_{d+1}$ in $d+1$ spacetime dimensions over a d-dimensional base
$X_d$
is given by
\eqn\cone{ds^2=d\rho^2+\rho^2 g_{ij}dx^i dx^j}
The cone $\cc_{d+1}$ has the property that the vector $\partial /
\partial \rho$ is conformally killing. The metric $g_{ij}; \,i,j=1 \dots
d$
determines the geometry of the base. $\cc_{d+1}$ is Ricci-flat iff $X_d$
is
Einstein with cosmological constant $(d-1)$ and is irregular at $\rho=0$
unless
$X_d=S^d$.

Our situation corresponds to d=5 and we will consider the base to be a $T^{1,1}$
space. A $T^{1,1}$ space is a particular example of $T^{p,q}$ spaces \pp. These
can be thought of as $U(1)$ fibrations over $S^2 \times S^2$. Let $0 \leq
\phi_i
\leq 2\pi \,$,$0
\leq
\theta_i \leq \pi
\,$ $,i=1,2$ parametrize the two $S^2$ and let $0 \leq \psi_i \leq 4\pi
\,$ be the
coordinate on the $U(1)$ fiber. The line element of $T^{p,q}$ is then given by
\eqn\tpq{\eqalign{ds^2=&\lambda_1(d\psi+pcos\theta_1 d\phi_1+qcos\theta_2
d\phi_2)^2
\cr
&+\lambda_2(sin^2\theta_1 d\phi_1^2+d\theta_1^2)+\lambda_3(sin^2\theta_2
d\phi_2^2 +d\theta_2^2) \cr}}
where the first term on the rhs is the vertical displacement along the fibre and
the other two terms are the line elements on the $S^2$'s. By ``forgetting'' one
of the $S^2$'s the $T^{1,1}$ space can be thought of as an $S^3$ fibration over
$S^2$ -the base being the $S^2$ we ``forget'' and the fibre being a $U(1)$
fibration over the other $S^2$. This fibration is actually trivial \foot{There
are various way to see this. See for example \kw.} and therefore $T^{1,1}$ is
topologically $S^2 \times S^3$.

If the following algebraic conditions are met
\eqn\rfcond{\eqalign{\Lambda \lambda_1&= {p^2 \over 2} ({\lambda_1 \over
\lambda_2})^2
+{q^2 \over 2} ({\lambda_1 \over \lambda_3})^2\cr &= ({\lambda_1 \over
\lambda_2})-{p^2 \over 2} ({\lambda_1 \over \lambda_2})^2\cr
&= ({\lambda_1 \over \lambda_3})-{q^2 \over 2} ({\lambda_1 \over
\lambda_3})^2\cr}}
equation \tpq\ describes an Einstein manifold of cosmological constant
$\Lambda$.

In the case $p=q=1$ and $\Lambda=4$ the cone over $T^{1,1}$ is Ricci-flat and
\rfcond\ implies
\eqn\rfcondft{\lambda_1={1 \over 9}; \,\,\,\,\lambda_2=\lambda_3={1 \over
6} }

The spaces $T^{p,q}$ also have a coset description as $SU(2) \times SU(2)/U(1)$,
which is another way to see the $S^2 \times S^3$ topology. It will pay off to
make a digression on the geometry of coset spaces which will eventually help us
give a useful description of the deformed conifold. For a more comprehensive
account one should consult the literature \refs{\cs, \cdf}.

\subsec{The geometry of coset spaces.}

In this section we use techniques of coset spaces to give a generalization of
the metric  \tpq. This generalized metric for $T^{1,1}$ will appear naturally in
the following when we discuss the deformed conifold.

Consider a Lie group $G$ and a subgroup $H \in G$ generated by $\{ H_i; \,
i=1
\dots dimH \}$ such that $G$ is generated by $\{ H_i, E_a; \, a=1
\dots dimG-dimH \}$ and
\eqn\redal{\eqalign{[H_i,H_j]&=c_{ij}{}^k H_k; \cr
[H_i,E_a]&=c_{ia}{}^b E_b;  \cr [E_a,E_b]&=c_{ab}{}^d E_d+c_{ab}{}^i H_i \cr}}
We call {\it left cosets} the elements of the form $g.H, \, g \in G$. To
parametrize the coset we choose a particular group element in each coset
which
we call the {\it coset representative} $L(\phi^\alpha)$. Here $\{
\phi^\alpha,
\alpha =1 \dots dimG-dimH \}$ are coordinates on $G/H$. The element
$L^{-1}
\partial_\alpha L$, where $\partial_\alpha:=\partial / \partial
\phi^\a$, is in the Lie algebra of $G$. The expansion
\eqn\vb{L^{-1} \partial_\alpha L=e^a_\a E_a +\omega_\a^i H_i}
defines the vielbein $e^a_\a(\phi)$ and the {\it H-connections}
$\omega_\a^i(\phi)$. Coset manifolds $G/H$ have at least an isometry $G'
\times
N(H)/H$, where the latter factor is the normalizer (the largest
subgroup of G of which H is a normal subgroup) and $G=G' \times
(U(1)$ factors common to $G$ and to $N(H)/H)$.

A metric on $G/H$ preserving the isometries is given by
\eqn\ctmc{g_{\a \b}= h_{ab} e^a_\a e^b_\b}
where $h_{ab}$ is an H-invariant tensor
\eqn\hinv{c_{ia}{}^c h_{cb}+c_{ib}{}^c h_{ac}=0}
As an application, we can reproduce the metric \tpq\ for $p=q=1$ as
follows. Parametrizing using Euler angles the group element of
$SU(2) \times SU(2)$ can be written as
\eqn\repfsut{e^{i\sigma_3 \phi_1/2}e^{i\sigma_2 \theta_1/2}e^{i\sigma'_3
\phi_2/2}
e^{i\sigma'_2 \theta_2/2}e^{i(\sigma_3 + \sigma'_3)(\psi_1
+\psi_2)/4}e^{i(\sigma_3 -\sigma'_3)(\psi_1 -\psi_2)/4} }
where $\{ \sigma_i \}$ and $\{ \sigma'_i \}$ obey the algebra of the Pauli
matrices with $[\sigma,\sigma']=0$. We take the coset representative to be
\eqn\csrep{L=e^{i\sigma_3 \phi_1/2}e^{i\sigma_2 \theta_1/2}e^{i\sigma'_3
\phi_2/2}
e^{i\sigma'_2 \theta_2/2}e^{i(\sigma_3 + \sigma'_3)\psi/2}}
Let
\eqn\bsdf{\eqalign{E_{1,2}={i \over 2}\sigma_{1,2}&; \,\,\,
 E_{3,4}={i \over 2}\sigma'_{1,2} \cr
E_5={i \over 4}(\sigma_3+\sigma'_3)&; \,\,\, H={i \over
4}(\sigma_3-\sigma'_3)
\cr}}
From
\vb\ we get
\eqn\expvb{\eqalign{e^1 &=-sin\theta_1 d\phi_1 \cr e^2&=d\theta_1 \cr
e^3&=cos\psi sin\theta_2 d\phi_2-sin\psi d\theta_2 \cr e^4&=sin\psi
sin\theta_2
d\phi_2+cos\psi d\theta_2 \cr e^5&=d\psi+cos\theta_1 d\phi_1 +cos\theta_2
d\phi_2 \cr}}
Since the coset representation $c_{ia}{}^b$ of $H$ (cf. \redal) is block
diagonal
\eqn\cstrep{c_{ia}{}^b={i \over 2}\pmatrix{
-\sigma_2 & 0 & 0 \cr 0 & \sigma_2 & 0 \cr 0 & 0 & 0 \cr}}
the most general H-invariant tensor $ h_{ab}$ is of the form
\eqn\genhinvt{ h_{ab}=\pmatrix{
A\I1_2 & B\I1_2 & 0 \cr B\I1_2 & C\I1_2 & 0 \cr 0 & 0 & D \cr}}
Substituting to \ctmc\ we get the $T^{1,1}$ metric
\eqn\gentoomt{\eqalign{ds^2=D&(e^5)^2 +A((e^1)^2+(e^2)^2)
+C((e^3)^2+(e^4)^2)+2B(e^1e^3+e^2e^4) \cr =D&(d\psi+cos\theta_1
d\phi_1+cos\theta_2 d\phi_2)^2 +A(sin^2\theta_1
d\phi_1^2+d\theta_1^2)+C(sin^2\theta_2 d\phi_2^2 +d\theta_2^2)\cr
&+2B[cos\psi(d\theta_1 d\theta_2 -d\phi_1 d\phi_2 sin\theta_1 sin\theta_2)
+sin\psi(sin\theta_1 d\phi_1 d\theta_2 +sin\theta_2 d\phi_2 d\theta_1)]\cr}}
Note that for $B
\neq 0$ the metric above cannot be
Einstein. Indeed one finds for the $a=2, \, b=4$ component of the Ricci
tensor
$R_{ab}$
\eqn\neqvofm{R_{24}= {-B^2(cos2 \theta_1 - cos2 \theta_2) csc^2 \theta_1
          csc^2 \theta_2 \over 8(A-B)^{3/2}
          (A+B)^{3/2}}}
For $B=0$ the metric reduces to the standard metric
\tpq\ with $p=q=1$.

For simplicity we will consider the case $A=C$. It is useful to make a
redefinition
\eqn\nvlbs{e^a \rightarrow g^a:= P^a{}_b e^b; \,\,\,\, P:=\pmatrix{
{1 \over \sqrt{2}}\I1_2 & -{1 \over \sqrt{2}}\I1_2 & 0 \cr {1 \over
\sqrt{2}}\I1_2 & {1 \over \sqrt{2}}\I1_2 & 0
\cr 0 & 0 & 1 \cr}}
In this basis \gentoomt\ becomes diagonal
\eqn\gendgmt{ds^2= D(g^5)^2+(A+B)[(g^3)^2+(g^4)^2]+(A-B)[(g^1)^2+(g^2)^2]}
We remark that the first two terms describe a squashed $S^3$. Indeed if we
define
\eqn\defoft{T:=L_1 \sigma_1 L_2^{\dagger}\sigma_1 }
we find
\eqn\rmtce{d\Omega_3^2:={1 \over 2}Tr(dT^{\dagger}dT)
= {1 \over 2}(g^5)^2+(g^3)^2+(g^4)^2}
Since $T$ is an $SU(2)$ matrix, the metric above is the standard round sphere
metric. Comparing with \gendgmt\ we conclude that the first two terms are the
line element of a (squashed) $S^3$. Moreover the last two terms
\eqn\tltt{ds_2^2:= (g^1)^2+(g^2)^2}
should describe a surface which is topologically an $S^2$ fibered
over the squashed $S^3$.

\subsec{The Ricci-flat cone over $T^{1,1}$ as a K\"{a}hler manifold.}

Before we come to the description of the deformed conifold let us
summarize some
of the results of
\co. In particular we will see how the Ricci-flat cone over $T^{1,1}$ can
be
thought of as a singular, non compact Calabi-Yau threefold.

Consider the cone in $\IC^4$ given by
\eqn\cnfld{\sum_{A=1}^4 (w^A)^2 =0}
Let us define a radial coordinate $\rho$ by
\eqn\radcoord{\rho^2=trWW^{\dagger} }
where $W:={1 \over \sqrt[2]}(w^i \sigma_i+iw^4\I1_2)$. The base of the
cone is
described by
\eqn\basecn{detW=0; \,\,\,\, \rho^2=constant}
A K\"{a}hler metric deriving from an $SU(2)
\times SU(2)$-invariant K\"{a}hler potential $K(\rho^2)$ reads
\eqn\khlrmt{ds_{\cc}^2=\vert trW^{\dagger}dW \vert^2 K(\rho^2)''+
tr(dW^{\dagger}dW) K(\rho^2)'}
Ricci-flatness determines the K\"{a}hler potential to be proportional to
$\rho^{4/3}$.

Let us parametrize \basecn\ as
\eqn\gensol{W=\rho Z; \,\,\,\, Z=L_1.Z^{(0)}.L_2^{\dagger}}
where $L_i, \, i=1,2$ are $SU(2)$ matrices. In terms of Euler angles
\eqn\sutm{L_j=\pmatrix{
 cos{\theta_j \over 2}e^{i(\psi_j+\phi_j)/2} & -sin{\theta_j \over
2}e^{-i(\psi_j-\phi_j)/2}\cr
 sin{\theta_j \over 2}e^{i(\psi_j-\phi_j)/2} &  cos{\theta_j \over
2}e^{-i(\psi_j+\phi_j)/2} \cr }}
and
\eqn\znaught{Z^{(0)}={1 \over 2}(\sigma_1 +i \sigma_2) }
Substituting to \khlrmt\ (after a redefinition of the radial coordinate
and by
setting $\psi:=\psi_1+\psi_2$) we find
\eqn\thcnovtoo{ds_{\cc}^2= (d\rho)^2 +ds_X^2}
with $ds^2_X$ the $T^{1,1}$ metric previously given in \tpq, for the unique
choice of constants $\lambda_1=1/9 \,$, $\lambda_2=\lambda_3=1/6$ which give a
Ricci-flat metric on the cone. The fact that the base is $T^{1,1}$ can also be
seen directly from \gensol\ by noting that there is a transitive $SU(2)
\times
SU(2)$ action with a $U(1)$ stabilizer embedded symmetrically in
the two $SU(2)$ factors
\eqn\stblen{L_1 \rightarrow L_1U; \,\,\, L_2 \rightarrow L_2 U^{\dagger};
\,\,\,\, U:=\pmatrix{ e^{i\theta} & 0\cr
 0 &  e^{-i\theta} \cr } \in U(1)}

\subsec{The deformed conifold: smooth noncompact $CY_3$}

In this section we will describe the deformation of the conifold.
The apex is replaced by an $S^3$. Insisting on Ricci-flatness we
get a smooth noncompact $CY_3$. The $\rho =constant \neq
\varepsilon$ surfaces are still $T^{1,1}$ spaces whose geometry is
described by the generalized metric \gentoomt. We also examine the geometry near
the apex.

Consider deforming \cnfld\ to
\eqn\defcnfld{detW=-\varepsilon^2/2}
We can again define a radial coordinate as in \radcoord\ but now $\rho$ is
bounded below by $\varepsilon$. We can parametrize the $\rho=constant$
surfaces
by
\eqn\defgensol{W_{\varepsilon}=\rho Z_{\varepsilon}; \,\,\,\,
 Z_{\varepsilon}=L_1.Z^{(0)}_{\varepsilon}.L_2^{\dagger}}
where $L_i$'s are as before and
\eqn\zneps{\eqalign{Z^{(0)}_{\varepsilon}&=\pmatrix{
 0 & a    \cr
 b &  0 \cr }; \cr
 a={1 \over 2}(\sqrt{1+{\epsilon^2 \over \rho^2}}&+ \sqrt{1-{\epsilon^2
\over \rho^2}});
  \,\,\,\, b= {\epsilon^2 \over 2\rho^2}a^{-1}\cr}}
For $\rho \neq \epsilon$ there is again a transitive $SU(2) \times SU(2)$
action
with a $U(1)$ stabilizer. For $\rho = \epsilon$ however, the stabilizer is
enhanced to the whole of SU(2)
\eqn\stblenh{L_1 \rightarrow L_1U; \,\,\, L_2 \rightarrow L_2 \sigma_1 U
\sigma_1,}
where $U \in SU(2)$. The surfaces $\rho=constant$ are again $T^{1,1}$ spaces
except for the surface $\rho = \epsilon$ which is an $S^3$.

The metric is obtained by substituting in \khlrmt\ $W$ given by
\defgensol. The result is \foot{In the original
version of this paper this equation had an error. We would like to
thank I. Klebanov and M. Strassler for pointing it out and also
for pointing out the errors it propagated to the next sections. }
\eqn\dfmt{\eqalign{ds^2=[(\rho^2 \gamma' -\gamma)(1-{\varepsilon^4
\over \rho^4})&+\gamma ]\left( (d\rho)^2 {1 \over
\rho^2(1-{\varepsilon^4\over \rho^4})} +{1 \over 4}
(d\psi+cos\theta_1 d\phi_1+cos\theta_2 d\phi_2)^2\right) \cr &+
{\gamma \over 4}(sin^2\theta_1 d\phi_1^2+d\theta_1^2+sin^2\theta_2
d\phi_2^2 +d\theta_2^2) \cr +\gamma {\varepsilon^2 \over 2\rho^2}
[ cos\psi (d\theta_1 d\theta_2 -&d\phi_1 d\phi_2 sin\theta_1
sin\theta_2) +sin\psi(sin\theta_1 d\phi_1 d\theta_2 +sin\theta_2
d\phi_2 d\theta_1) ] \cr}}
where $\gamma:=\rho^2 K'(\rho^2)$ and $\gamma':=\gamma'(\rho^2)$. The metric on
the $\rho=constant$ slice is of the generalized form
\gentoomt.

Requiring Ricci-flatness and correct asymptotic behaviour of the metric
\dfmt\ leads to the following differential equation
\eqn\rfgm{\rho^2(\rho^4-\varepsilon^4)(\gamma^3)'+3\varepsilon^4
\gamma^3-2\rho^8=0}
The general solution reads
\eqn\gensolg{\gamma=
(c+{\varepsilon^4 \over 2}(sinh2\tau-2\tau))^{1/3}(tanh\tau)^{-1}}
where $c$ is a constant. In the limit $\rho / \varepsilon \rightarrow
\infty$
the constant $c$ can be dropped and the solution asymptotes its ``cone''
value.
For $c \neq 0$, $\gamma$ diverges at $\rho=\varepsilon$. From now on we
set $c$
to zero.

\subsec{The near-horizon limit $\rho \rightarrow \varepsilon$}

Let us examine in more detail the limit $\rho \rightarrow \varepsilon$. We
make
a change of variables
\eqn\chvr{\delta=\rho-\varepsilon}
so that
\eqn\lmd{0 \leq \delta < \infty}
We find that the metric \dfmt\ takes the form:
\eqn\lnnhlev{ds^2 \sim R_{\varepsilon}^2[(d\upsilon)^2+
d\Omega_3^2 + {\upsilon^2 \over 2} ds_2^2]  }
where
\eqn\fdofu{\upsilon:= \sqrt{ {2\delta \over \varepsilon}}
\rightarrow 0}
and
\eqn\defofr{R_{\varepsilon}:={1 \over \sqrt{2}}({2\varepsilon^4 \over
3})^{1/6} }
is the radius of the $S^3$ on the apex, $d\Omega^2_3$ is the round
sphere element defined in
\rmtce\ and $ds_2^2$ was defined in \tltt\ and describes a fibre of
$S^2$
topology.

\newsec{D3 on $S^3$}

We will now consider the near-horizon limit in the ten-dimensional sense
$r
\sim
0; \,\,\, \rho \sim \varepsilon$. The ten-dimensional metric in the absence of
D3 is of the form \mtritapd, and our solution-generating method applies, only
near the $S^3$ at the apex of the deformed conifold. We therefore want to
emphasize that we only have a metric (implicitly) for the geometry near the
horizon. Presumably there is a complete solution, corresponding to a D3 whose
near-horizon limit coincides with the one given here but we were not able to
obtain it. Of course the remark at the beginning of section 3 applies here as
well: as we do not have the charge distribution explicitly, nothing excludes the
possibility that this solution corresponds to a completely ``smeared'' D3.

We define a ten-dimensional radial coordinate $U$ and an angle $\theta$ by
\eqn\dfnou{r=Ucos\theta; \,\,\,\, \upsilon= {U \over
R_{\varepsilon}}sin\theta}
where $\upsilon$ was defined in \fdofu. Taking \lnnhlev\ into account, we see
that the metric is of the form
\cfv\ as advertised
\eqn\assadv{\eqalign{ds^2= &-dt^2 +R_{\varepsilon}^2d\Omega_3^2 \cr
 &+dU^2+ U^2\left( d\theta^2 +cos^2\theta d\Omega_2^2
 +{1 \over 2} sin^2\theta ds_2^2 \right)  \cr}}
where the first line contains the directions parallel to the D3 and the second
line contains the transverse geometry. As already remarked below \tltt, $ds_2^2$
is the line element of an $S^2$ fibred over an $S^3$. The vielbein $\tilde{e}^A$
of the 10d metric in the presence of D3 has to be compatible with the connection
given in
\cnanst\ as already explained. In particular if we define
\eqn\ighjk{h^A:=\tilde{e}^A-e^A }
we have
\eqn\cpkkjh{D(\omega)h^A+e^A \wedge dln\Delta_{(A)}+  e^A
\wedge h^B \partial_B ln\Delta_{(A)}+
e_B
\wedge h^B \partial^A ln\Delta_{(B)}=0}
The geometry in the presence of D3 is given by equations \cpkkjh,
\harmcnd,
but we will not be more explicit here. However the case of the shrinking cycle
limit $\varepsilon \rightarrow 0$ ($R_{\varepsilon} \rightarrow 0$) of the
near-horizon geometry and the case of the flat D3 limit can be analyzed
explicitly.

\subsec{The $\varepsilon \rightarrow 0$ limit of the near-horizon
geometry}

As we see from \lnnhlev, in the $\upsilon \rightarrow 0; \,\,  \varepsilon
\rightarrow 0$ ($R_{\varepsilon}
\rightarrow 0$) limit the near-horizon geometry becomes effectively
four-dimensional and is split orthogonally between the directions parallel
to
the brane (the time direction) and the transverse directions. Equation
\harmcnd\ reduces to the harmonic condition in three spatial dimensions
and we recover $AdS_2 \times S^2$ as usual.

\subsec{The flat D3 limit.}

We can examine the limit where the D3 is seen as flat, i.e. for
$R_{\varepsilon} \rightarrow \infty$ with $U$ fixed. All
curvatures vanish in this limit. Moreover, for $\delta \rightarrow
\infty$ we are moving away from the brane and we expect to recover
ten dimensional Minkowski space as a solution. Indeed in this case
we easily see that $\Delta \rightarrow constant$.

\newsec{M-branes, $M^{pq}$ spaces and T-duality}

We conclude with a brief discussion on higher-dimensional conifolds. Indeed one
can also consider conifolds with seven-dimensional bases in $M$-theory
- as
has been done extensively in the eighties in compactifications of the
eleven-dimensional supergravity \dpn\ but also very recently. We shall
concentrate only on the example that is directly related to our previous
discussion, namely $M^{10}$ (and its ``T-dual" $M^{01}$). Just as $T^{p,q}$
these are constructed as $U(1)$ bundles \refs{\wittkk, \crw}:
\eqn\mpr{M^{pq}= M^{pq0} = {S^5 \times S^3 \over U(1)}.}
Since odd spheres can be thought of as $U(1)$ bundles over projective spaces,
the factoring  leads to identification of the two fibers in \mpr\ and as a
result $M^{pq}$ can be thought as a $U(1)$ bundle over $\IC\IP^2 \times S^2$
with the topology depending on the ratio of $p$ and $q$ (homotopically all these
spaces are the same). While in general the isometry group is $SU(3)
\times SU(2) \times U(1),$ we are interested in cases with larger symmetry
-
$SO(6) \times SO(3)$ for $M^{10} = S^5 \times S^2$ and $SU(3)
\times SU(2) \times SU(2),$ for $M^{01}=\IC\IP^2 \times S^3.$

A cone over $M^{10}$, ${\cal C}(M^{10})$, can be almost everywhere described by
a quadric in $\IC^6$ with two real planes intersecting it. Trying to resolve the
singularity, we end up replacing the apex of the cone either by $S^2$ factor of
by $S^5$. Differently from the previous case where only the deformation of
${\cal C}(T^{1,1})$ (finite size $S^3$) was of interest for us here we get a
``duality" between the factors
-
when putting $M$-theory on the cone, we can get a three-dimensional back
hole
either by wrapping $M2$ on $S^2$, or by the dual procedure of wrapping
$M5$ on
$S^5.$

Similarly, a two-dimensional black hole can be constructed by wrapping D3
on the
shrinking $S^3$ in ${\cal C}(M^{01}).$ A circle compactification of the
previous
case accompanied with $T$-duality should relate this to the $M2$ and $M5$
discussed above. Indeed, as known \dlp, $T$-duality untwists $U(1)$
bundles
interchanging the bases of the two cones. All these cases have no
supersymmetry
preserved since both $M^{10}$ and $M^{01}$ admit no Killing spinors.

\bigskip
\centerline{\bf Acknowledgments}\nobreak
\medskip

We would like to thank Gregory Moore for suggesting to us the
problem of D3 on a conifold as a more tractable example of a brane
on a non-trivially embedded cycle, for suggestions on a
preliminary version of this paper and for numerous useful
discussions. Helpful discussions with V. Moncrief, G. Zuckerman
are also gratefully acknowledged. This work is supported by DOE
grant DE-FG02-92ER40704.

\listrefs
\end